\documentclass[preprint,journal]{vgtc}       

\usepackage{mathptmx}
\usepackage{graphicx}
\usepackage{times}

\usepackage{amsmath}

\usepackage[bookmarks,backref=true,linkcolor=black]{hyperref} 
\hypersetup{
  pdfauthor = {},
  pdftitle = {},
  pdfsubject = {},
  pdfkeywords = {},
  colorlinks=true,
  linkcolor= black,
  citecolor= black,
  pageanchor=true,
  urlcolor = black,
  plainpages = false,
  linktocpage
}

\onlineid{0}

\vgtccategory{Algorithm/Technique}

\vgtcinsertpkg

\preprinttext{Submitted for publication}

\title{3D Density Histograms for Criteria-driven Edge Bundling}

\author{Daniel C. Moura}
\authorfooter{
\item
 Daniel C. Moura is with Instituto de Telecomunica{\c c}\~oes, Departamento de Engenharia  Electrot\'ecnica e de Computadores, Faculdade de Engenharia, Universidade do Porto. E-mail: daniel.c.moura@gmail.com.
}

\shortauthortitle{Anonymous: {3D} Density Histograms for Criteria-driven Edge Bundling}

\abstract{This paper presents a graph bundling algorithm that agglomerates edges taking into account both spatial proximity as well as user-defined criteria in order to reveal patterns that were not perceivable with previous bundling techniques. Each edge belongs to a group that may either be an input of the problem or found by clustering one or more edge properties such as origin, destination, orientation, length or domain-specific properties. Bundling is driven by a stack of density maps, with each map capturing both the edge density of a given group as well as interactions with edges from other groups. Density maps are efficiently calculated by smoothing 2D histograms of edge occurrence using repeated averaging filters based on integral images.

A CPU implementation of the algorithm is tested on several graphs, and different grouping criteria are used to illustrate how the proposed technique can render different visualizations of the same data. Bundling performance is much higher than on previous approaches, being particularly noticeable on large graphs, with millions of edges being bundled in seconds.
} 

\keywords{Graphs, networks, visualization, imaged-based bundling, edge clustering}

   \teaser{
   \centering
   \includegraphics[width=16cm]{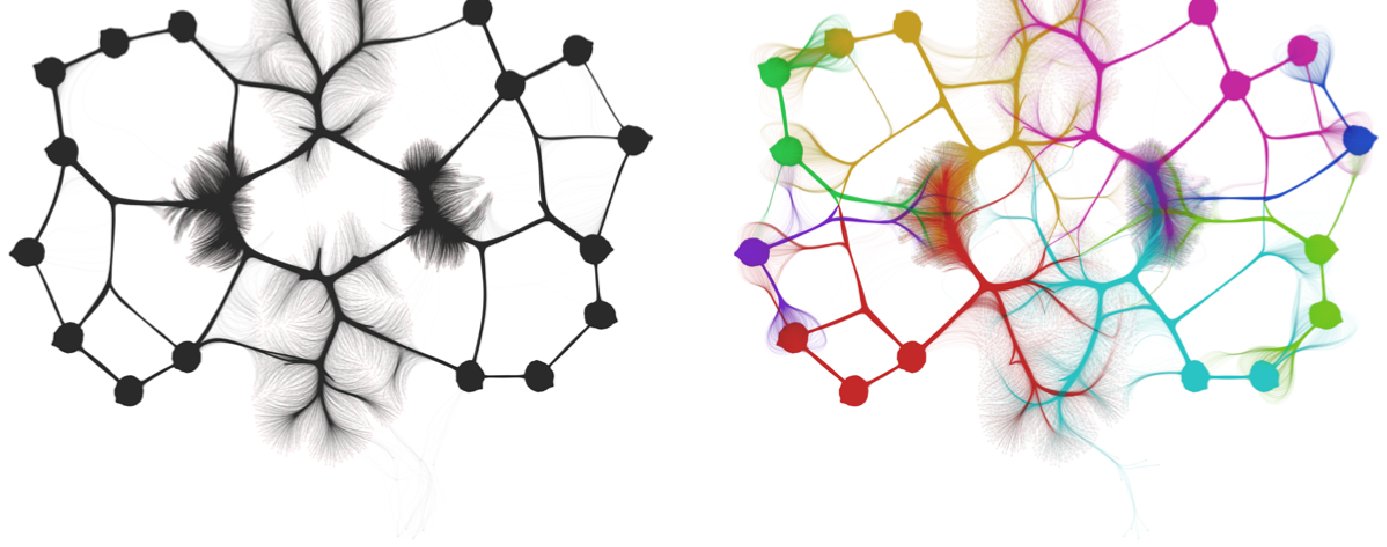}
   \caption{Bundling of a graph with 5 million edges (\texttt{pattern1} from the University of Florida Sparse Matrix Collection \cite{matrixcoll}) using the proposed algorithm. The left drawing shows the result of using only spatial proximity as bundling criterium, while the right uses properties of the edges (namely, their origin and destination coordinates) clustered in 8 groups to show patterns that were not perceivable in the left drawing. Bundling times of 33 and 94 seconds respectively with a CPU implementation on a Macbook Air. }
  }

\begin{document}


\firstsection{Introduction}

\maketitle



Graphs are one of the most used data structures for modeling relations between objects in many application domains, such as, transportation, biology, computer science and social networks. Node-link diagrams \cite{gd1999graph}, in which nodes are drawn as points and edges as straight lines, are commonly used to visualize graphs. However, as the size of graphs increase, visualization becomes more and more cluttered, making difficult to understand the graph structure. 

Edge bundling methods have gained increasing attention in recent years due to their ability to visualize large graphs with an emphasis on graph structure. 
In edge bundling approaches, the input of the problem is a graph defining the positions of the nodes and their edges. When nodes positions are not available, a layout algorithm is used as pre-processing to distribute the nodes in space. The goal of edge bundling is to find the drawing paths of the edges that allow to aggregate similar edges, decreasing clutter and providing a better understanding of the graph connectivity patterns. Several bundling algorithms have been proposed in recent years \cite{GBEB, SBEB, MINGLE, HEB, FDEB,  KDEEB, WR, OB, DEB, IBEB}. However, most approaches to the problem either cannot coupe with graphs with millions of edges or require high computation times. Additionally, users typically have no control over the bundling criteria, which is normally limited to spatial proximity. 

In this paper, we propose a novel approach to construct edge bundles of general graphs. First, we propose a new image-based pipeline for efficiently attracting edges to areas of the drawing with higher edge density. While our approach is based on the basic principle of edge density proposed in \cite{KDEEB}, instead of using a classical Kernel Density Estimation, we propose estimating density in two steps: a) calculating a 2D histogram of edge frequency followed by b) smoothing the histogram with a fast filter.  The computational complexity of the second step does not depend on the size of the graph and, thus, it allows higher computational efficiency on large graphs.  Edge rerouting is then performed based on the gradient of the density map with an optimized step that improves convergence by taking into account the local distribution of edge density. Moreover, we propose a framework that allows driving the edge bundling process in a way that user-defined edge properties (either spatial or not) are taken into account. This is achieved by the use of 3D density histograms that comprise multiple density maps to guide different groups of edges taking into account interactions between groups.  When rendering, by making edges’ color dependent of their group, one may reveal patterns of node connectivity that are not perceivable with previous edge bundling algorithms. In summary, the main contributions of this work are the following:
\begin{itemize}
\item A new and much more scalable pipeline that allows to decrease computing times up to 2 orders of magnitude in graphs composed by millions of edges;
\item A new framework that allows to bundle edges based on user-defined criteria, revealing patterns that were not perceivable with previous techniques.
\end{itemize}

The reminder of the paper is organized as follows. Section \ref{sec2} presents related work on edge bundling. Section \ref{sec3} proposes our bundling algorithm. Section \ref{sec4} provides implementation details. Section \ref{sec5} presents results of our algorithm, and compares and discusses results with related work. Section \ref{sec6} concludes the paper and identifies future research directions.

\section{Related work}
\label{sec2}

Edge bundling algorithms try to create uncluttered drawings of graphs by routing related edges so that they share similar paths. It is often defined as a sharpening operation over a graph that concentrates edges in areas with high edge density, promoting white spaces in areas with low edge density and, thus,  improving graph readability.

Holten coined the term edge bundling when he proposed Hierarchical Edge Bundling (HEB) for compound (hierarchy-and-association) graphs \cite{HEB}. HEB bundles edges using B-splines, following the control points defined by the hierarchy. Gansner and Koren bundle edges in a circular node layouts by merging edges so that the resulting splines share some control points while minimizing the total amount of ink needed to draw the edges \cite{gansner2007circularlayouts}. 

Cui et al. proposed Geometry Based Edge Bundling (GBEB) \cite{GBEB}, one of the first methods suitable for bundling general undirected layouts. Bundling is performed by forcing edges to pass through the same control points of a control mesh.  Holten and van Wijk used principles of physics to attract edges that are close to each other in their Force-directed edge bundling (FDEB) algorithm \cite{FDEB}. Bundled graphs were considered to be smoother and easier to read than on previous approaches, but the computational complexity of the algorithm is high, making it slower than GBEB. FDEB was later extended to separate opposite-direction bundles in directed graphs in the Divided Edge Bundling (DEB) algorithm \cite{DEB}. Control meshes were revisited in the Wind Roads (WR) algorithm proposed by Lambert et al. \cite{WR}. In WR, graph edges are routed along mesh edges using a shortest path algorithm. The  computational performance of the algorithm was better than FDEB and comparable to GBEB. Later, Lambert et al. extended the algorithm to 3D \cite{WR3D}. 

In 2011, Gansner et al. proposed a multilevel agglomerative edge bundling method (MINGLE) based on minimizing the ink needed to represent edges, with additional constraints on the curvature of the resulting splines \cite{MINGLE}. While drawings are more cluttered and less smooth than some of the previous techniques, MINGLE remains the fastest algorithm and the only that demonstrated being able to coupe with graphs comprised by millions of edges, although requiring several minutes to process them. Pupyrev et al. proposed Ordered Bundles (OB) \cite{OB}, where edge routing is accomplished through a heuristic that tries to minimize the total length of the paths together with their ink. After bundling, the method separates edges belonging to the same bundle to allow detailed local views. Computing times are higher than several of the previous approaches but, in the other hand, this technique offers a level of detail not previously achieved. 

In \cite{IBEB}, an image-based approach to edge bundling was proposed that was subsequently extended to create the Skeleton-based edge bundling (SBEB) algorithm \cite{SBEB}. SBEB uses skeletons of edges that are similar in terms of position information. Edges similarity is found by clustering using a specific similarity metric. In 2012, a new image-based approach was proposed based on the principle of edge density. The algorithm, Kernel Density Estimation Edge-Bundling (KDEEB), attracts edges into areas of the drawing with higher edge density, promoting white spaces where edge density is lower and, thus, increasing graph readability. KDEEB generates smooth uncluttered bundles in a fraction of the time of previous algorithms, with the exception o MINGLE that remains fastest.  Since the algorithm proposed in this paper follows the same principle as KDEEB, this method is explained next in more detail.

\subsection{Graph bundling based on edge density}

KDEEB is based on creating a density map with the size of the drawing that is used to guide the bundling process. The density map is created by Kernel Density Estimation (KDE) \cite{kde}. The authors' implementation involves splatting a kernel on the density map several times along each edge. 
The kernel size, $h$, defines the influence of the neighbor edges and, in KDE terminology, it is known as kernel bandwidth.  Edges are attracted to higher density areas by moving them in the direction of the gradient of the density map. For routing edges while keeping their endpoints fixed, edges are represented by a set of sample points $x_{ij}$ equally spaced by a sampling step $\delta$. After calculating the density map, a process denominated as advection translates each point $x_{ij}$ by a vector with the direction of the gradient in the density map at that point and magnitude equal to the kernel bandwidth. Finally, a Laplacian smoothing is applied to all points $x_{ij}$ to eliminate artifacts caused by edge advection and by the discretized nature of the density map. This process is repeated iteratively for a fixed number of iterations (8..10). At the beginning of each iterations, points $x_{ij}$ are resampled to ensure an uniform density of points along the edge. For guaranteeing convergence, the kernel bandwidth $h$ is decreased at each step $i$ by a geometric series $h_i =\lambda^i h_{max}$, where $h_{max}$ is the initial kernel bandwidth, and $\lambda$ is a reduction factor (between 0.5 and 0.9). Thus, as the number of iterations increases the neighborhood decreases as well as the advection of the sample points.

\section{Edge Bundling based on density histograms}
\label{sec3}

Our approach follows the same basic principle of KDEEB, which is attracting edges to the areas of the drawing with higher edge density. In KDEEB, density estimation involves splatting a kernel several times along each edge, making the process highly dependent of the size of the problem, with large graphs requiring a high number of splatting operations. In this paper, we propose estimating density in two steps: a) calculating a 2D histogram of edge frequency followed by b) smoothing the histogram with a fast filter in order to include neighborhood information and attenuate the effects of discretization. The computational complexity of the second step does not depend on the size of the graph, which makes the method more scalable. We also propose optimizing advection by taking into account the local distribution of edge density when translating the sample points. In an iteration of KDEEB, all points are equally translated even if they get placed in positions with lower density than on the original positions. This potentially delays convergence and makes the algorithm highly dependent of the kernel bandwidth, which defines the magnitude of the translation vector. 

Previous approaches to edge bundling aggregate edges mainly based on an edge similarity criteria that is algorithm-specific. We propose a framework that allows driving the edge bundling process in a way that user-defined edge properties are taken into account. For instance, one may choose to bundle edges that share similar orientation, destination, or length in order to have a higher insight of the relations between nodes.  For achieving this, edges are assigned to bins according to a given criteria and a third dimension is added to the edge density histogram, creating a stack of 2D histograms, with each layer being associated to a bin. Bundling is performed for each layer, while interactions between bins are also modeled, allowing to control attraction and repellence between edges based on the given criteria. Finally, when rendering, by making edges’ color dependent of their bin, one may reveal patterns of node connectivity that are not perceivable with previous edge bundling algorithms.

\begin{figure}[htb]
 \centering
 \includegraphics[width=1.0\columnwidth]{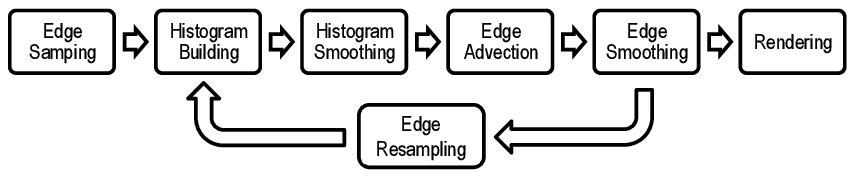}
 \caption{Pipeline of the proposed bundling algorithm. }
 \label{fig:flowchart}
\end{figure}

The pipeline of the proposed algorithm (Fig. \ref{fig:flowchart}) starts by sampling edges in equally spaced points ($x_{ij}$). When processing directed graphs and when the bundling criteria is related to flow direction, we translate all sample points (with the exception of the end points) by a fixed amount (typically 0.1\% to 0.5\% of the drawing size) in the direction of the normal of the edge on the sample point. This immediately separates edges with opposite directions, potentially easing convergence. Then, histograms of edge density are created and finally smoothened, which are then used to guide the advection of the sample points. Points are finally smoothed using Laplacian smoothing, and the process is repeated until convergence, starting by a edge resampling operation that creates and remove sample points when consecutive points are too far or too close, respectively. When the bundling is finished, the graph is finally rendered generating a 2D drawing. 

\subsection{2D Density Histograms}
Let us first consider the case of using a single spatial histogram to model edge density.  
Let $G = \{N,E\}$ be a graph with $|N|$ nodes and $|E|$ edges in a two-dimensional space. Each node $n$ has a position in 2D denoted $x_n$. Each edge $e$ is defined by the position of a source and a target node, as well as a weight,  $e = (x_s, x_t, w)$. 
Given a two-dimensional histogram $H$, let $H(u,v)$ be the sum of the weights of all the edges that intersect the discrete position $(u,v)$,
\begin{equation}
H(u,v) = \sum{w_e}, \forall e \ni (u,v) \text{.}
\end{equation}
High values in $H$ are found in positions where edge density is high. While $H$ describes edge density distribution in space, it does not take into account neighborhood and, thus, it suffers from aliasing problems. Neighborhood information is introduced by smoothing the histogram. This may me achieved by convolving $H$ with a Gaussian distribution function of zero mean and $\sigma$ standard deviation, delivering the density map $\rho$,
\begin{equation}
 \rho = H * G(0,\sigma) \text{.}
\end{equation} 
This is a close and fast simplification of kernel density estimation with a Gaussian kernel \cite{kde_conv} that is used in several applications that do not require exact density estimates (e.g. \cite{biagioni2012map_inference,liu2012mining_gps_traces}). The size of the neighborhood is defined by the standard deviation of the Gaussian kernel ($\sigma$), with higher values of $\sigma$ allowing to aggregate more edges and visualizing macro patterns, while lower values of $\sigma$ generate more detailed, less bundled drawings. 

However, convolving an image is a complex operation that depends on both the size of  $H$ and on the size of the convolution kernel and, thus, on the size of the neighborhood. 
For edge bundling purposes, the smoothing operation is not required to be highly accurate but it should provide smooth maps that allow calculating stable first derivatives, which are needed to drive bundling. In \cite{kovesi2010almost_gaussian}, Kovesi has demonstrated that high quality approximations of Gaussian smoothing may be efficiently obtained independently of the size of the neighborhood by repeated filtering with averaging filters. The process is iterative, requiring 3 to 6 iterations, depending on the desired quality of the approximation, and is speeded-up by using integral images \cite{viola2001robust} for calculating the multiple averaging filters. We have found 3 iterations to generate sufficiently good approximations for edge bundling purposes.
At the end, the computational complexity of the convolution operation only depends on the number of iterations and on the size of $H$, which are all constants. 

The main advantages of using a convolution-based approach to edge density estimation is that the calculation of $H$ is a very straightforward operation, equivalent to rendering lines in an accumulation buffer, while the smoothing of $H$ does not depend on the size of the problem, making the process more scalable. Thus, it is expected to achieve lower computation times than KDEEB, especially with large graphs.

\subsection{3D Density Histograms}

When using a single 2D density map, all edges are attracted by the same density function. However, edges may have distinct properties or may belong to different groups. Thus, it makes sense that advection takes into account edges’ properties so that similar edges are attracted and dissimilar edges are either ignored or repelled. Let us consider that each edge belongs to a group and, thus, edges may be distributed by a set of bins according to the group that they belong to. Groups may be provided as input, may result from the discretization of variables or may be found by clustering edges using one or more variables. Here, each group has its own density map that controls the advection of all edges within that group. This way, one may control bundling taking into account a given criteria that splits edges into disjoint groups. By stacking the 2D histograms of all groups, $H$ becomes a 3D histogram. This distinct feature names  the proposed algorithm as 3D Histogram-based Edge Bundling (3DHEB).   

If no interaction is modeled among edges from different bins, $H(u,v,l)$ would simply represent the sum of the weights of all edges that belong to bin $l$ and intersect the discrete position $(u,v)$. However, modeling interactions among groups potentially creates richer drawings and may be used to decrease cluttering by repelling edges from different groups. Bin interaction is modeled by an interaction matrix $M$, allowing every edge $e$ to interact with every layer $l$ of the 3D histogram $H$,
\begin{equation}
H(u,v,l) = \sum{w_e M(l,b_e)}, \forall e \ni (u,v) \text{,}
\end{equation}
with $M(l,b_e)$ being the weight of the interaction (or attraction) between layer $l$ of the histogram and the edge bin $b_e$. If $M$ is an identity matrix the result would be the same as bundling each layer independently. A simple interaction model may be defined by attracting edges that belong to the same bin and repelling edges that belong to different bins,
\begin{equation}
M(l,b_e) = \left\{ 
  \begin{array}{c l}
    1 & \quad \text{if $l=b_e$}\\
     -\alpha/(B-1)  & \quad \text{if $l\neq b_e$}
  \end{array} \right.
\end{equation}
with $\alpha$ being a parameter controlling the amount of repellence and B being the number of bins of the histogram. In all drawings in this paper we have used $\alpha=0.25$. More complex models may be defined and the nature of the bins (e.g. ordinal, modulus, nominal) may be taken into account when defining $M$. 

After generating the 3D histogram, smoothing is applied independently to each layer.

\subsection{Edge Advection and Smoothing}
\begin{figure}[htb]
 \centering
 \includegraphics[width=0.9\columnwidth]{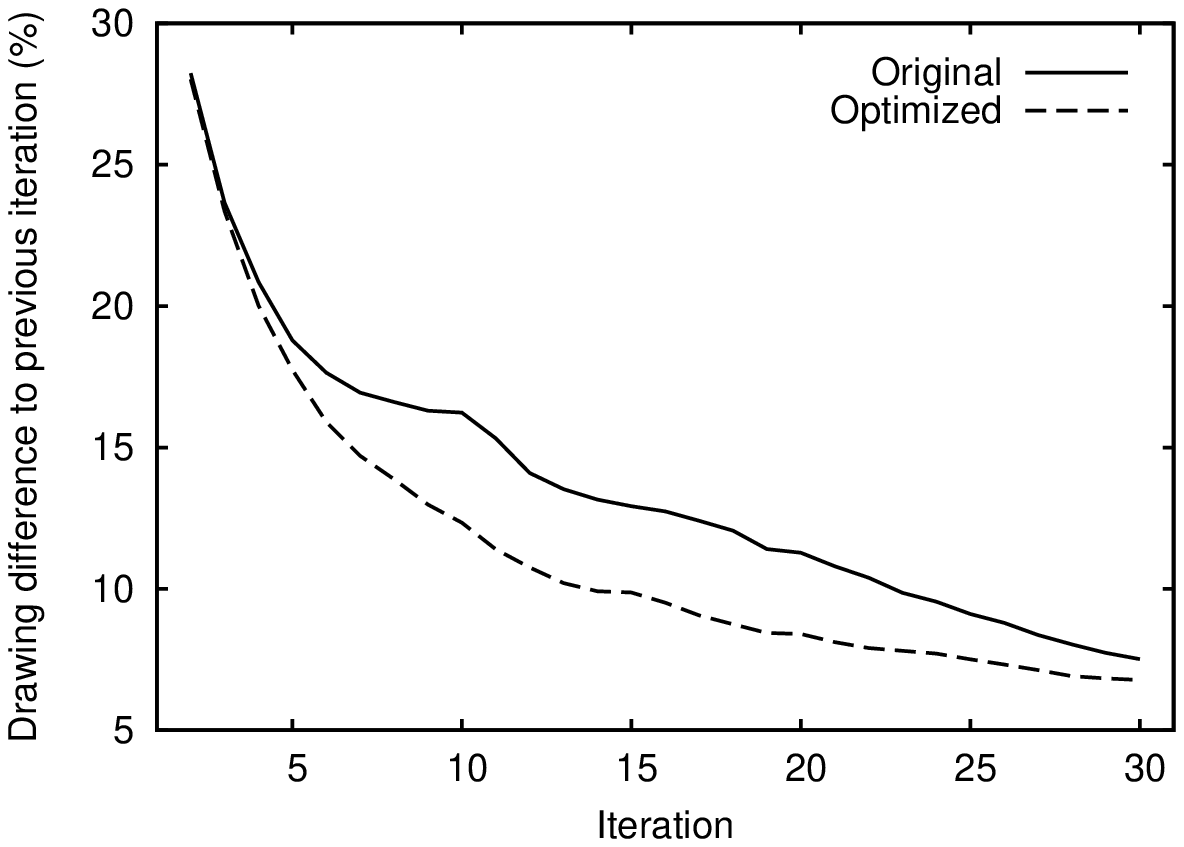}
 \caption{Comparison of the original advection step (which does not evaluate the density at the new location) with the proposed optimized step (which guarantees a higher density at the new location). (dataset: US migrations; $\lambda=0.9$; $h_{max} = 2\sigma)$. }
 \label{fig:diff2blind}
\end{figure}

Advection of the control points $x_{ij}$ of each edge $e_i$ is done by only taking into account the layer of the density map related to the edge's bin $b_e$. Like in \cite{KDEEB}, advection is done in the direction of the gradient of the density map, while taking into account the current kernel bandwidth $h$, 
\begin{equation}
\frac{dx}{dt} = \frac{h(t)\nabla\rho(t)}{\text{max}( \| \nabla\rho(t) \|, \epsilon)},
\end{equation} 
where $\nabla\rho$ represents the gradient of the density map and $\epsilon$ is a very small value to avoid divisions by zero. However, there is a substantial difference between our approach and KDEEB. Assuming a fixed magnitude $m = h(t)$ of the translation vector may move $x_{ij}$ to a place in the density map where the density is actually lower. This may happen, for instance, when $x_{ij}$ is already close to a local maximum on the density map. Thus, with the aim of improving convergence, we propose adjusting the magnitude of the translation vector so that it is guaranteed that the density in the new position is not lower than in the original position. This is done by, for each point $x_{ij}$, repeatedly decreasing $m$ (by half) until the new position is in a more dense area, or until $m$ is zero. By starting with the largest value of $m$ and decreasing it until an improvement is verified, we keep the exploratory nature of the KDEEB approach, while we guarantee that points are always translated to better positions. An experiment comparing the proposed approach with the original shows that convergence is improved (Fig. \ref{fig:diff2blind}). Thus, while the number of operations for a single iteration is increased, an equivalent bundling may be achieved with a lower number of iterations.

\begin{figure}[htb]
 \centering
 \includegraphics[width=1.0\columnwidth]{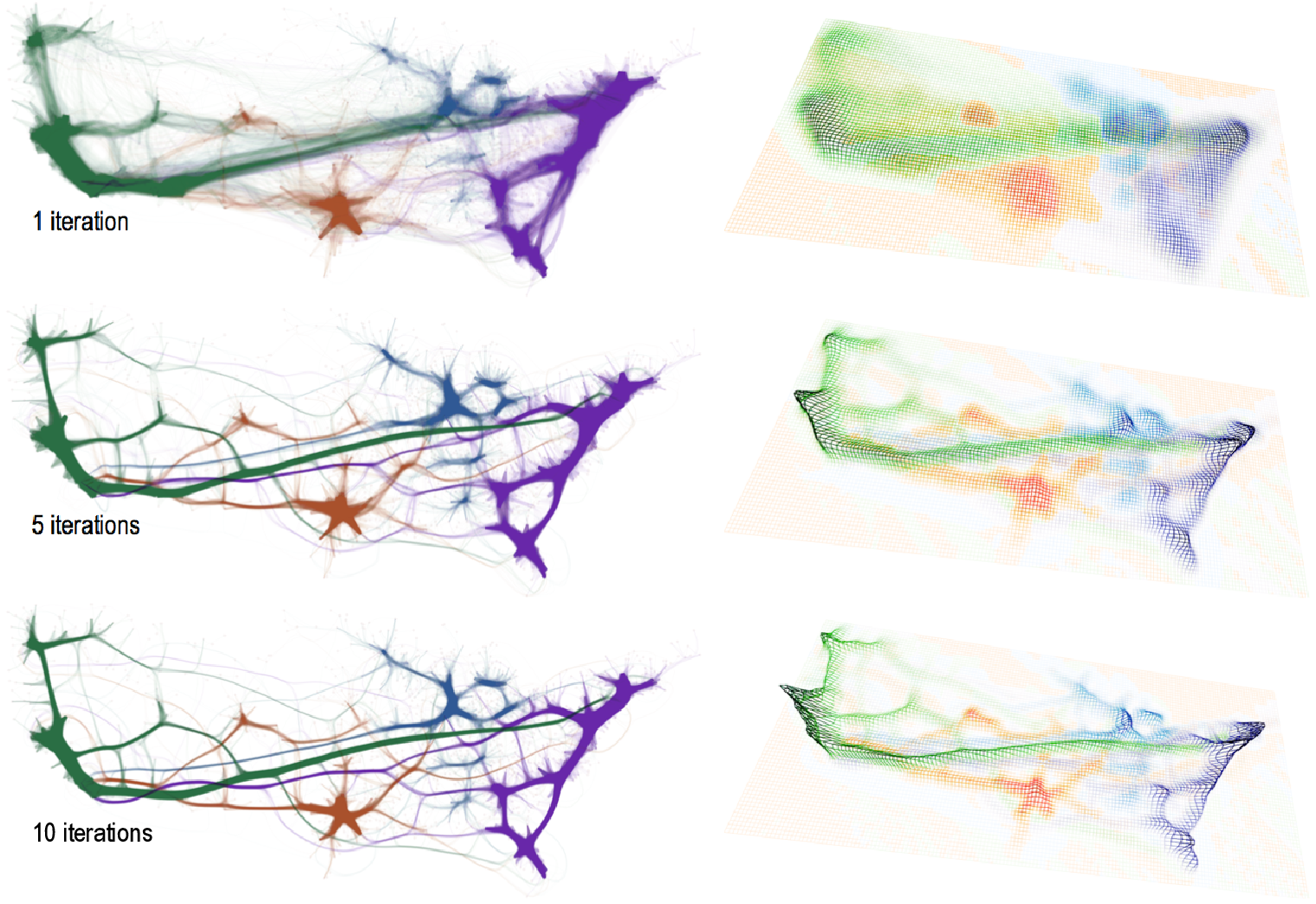}
 \caption{Partial results of the algorithm at different iterations (left column) for the US migration dataset and the correspondent density maps distinguished by color (right column). Edges are clustered in 4 groups according to the location of their target node. }
 \label{fig:sequence}
\end{figure}

In the first iteration, $h$ is equal to $h_{max}$, which in our work is defined as a function of $\sigma$, with typical values between $\sigma$ and $3\sigma$. In the following iterations $h$ is decreased by a $\lambda$ factor, which globally decreases the translation of $x_{ij}$ and guarantees that the bundling process stabilizes.  After advection, Laplacian smoothing is applied to all edges in a similar manner to \cite{FDEB,KDEEB}. Figure \ref{fig:sequence} illustrates the bundling process, where it is possible to observe that, as the number of iteration increases, the drawing becomes sharper and the number of overlaps between edges from different groups decreases. 

\subsection{Rendering}
Edges are rendered by drawing the polylines defined by their sample points $x_{ij}$. Our main concern when rendering the final drawing is highlighting patterns related to the bundling criteria. This is achieved by making the edge color dependent of its bin. Depending on the nature of bins, different color schemes are used. When bins are ordinal, a sequential color table is used to illustrate the relation between bins (e.g. darker colors for higher values). When bins are modal (e.g. direction), Hue coloring is used to guarantee a continuos color space. When bins are nominal, every bin has a distinct color.  Finally, if the bundling criteria aggregates edges based on flow direction, a sequential color table from blue to red may be optionally used to illustrate direction. An equivalent color scheme was used in \cite{DEB} where the authors where concerned with illustrating flow in graph networks, and its used was supported by \cite{holten2009_user_viz_edges}, which empirically demonstrated the use of color gradients as an effective way to communicate direction to viewers. 

Bundle weight is visually encoded by both alpha blending and line width. Alpha blending was explored several times in previous work (e.g. \cite{HEB,DEB,MINGLE,SBEB,KDEEB}) and makes bundles looking stronger when edge density is higher by accumulating the alpha of the edges that are drawn at the same places, while it enables to see unrelated edges that overlap. Still, in order to provide a better perception of the total bundle weight, we also encode bundle width as proposed in \cite{DEB}. At the end of the bundling process, bundled edges overlap and are highly condensed in the same positions, which may make the alpha blending by itself insufficient to highlight bundle weight. We propose using the edge density map to control the width of the line segments that make the edge polyline. When drawing an edge $e_i$, the line width between $x_{ij}$ and $x_{ij+1}$ is proportionally affected by the density on the sample points locations. This allows a fast width-based encoding that does not require to detected edge overlaps as in \cite{DEB}. Logarithmic scales may also be used when mapping density to width.

\section{Implementation details}
\label{sec4}

The bundling algorithm was implemented in the Processing 2.2.1 environment based on Java 7. Rendering is done with the Processing libraries, which use OpenGL 2.0 via JOGL. Processing is a simple cross-platform environment, allowing wide-spread use of the algorithm in the information visualization community and by other researchers that seek for a tool to visualize patterns in large graphs\footnote{With the publication of this article an open-source implementation of the algorithm in Processing will be made publicly available.}. 

The bundling algorithm was implemented for CPU and takes advantage of multi-threading. While Processing offers off-screen rendering supported by OpenGL, which could be used for generating histograms using the GPU, their precision is very low with only 8 bits per channel. Therefore, for the sake of simplicity and portability, we decided to make a multi-thread CPU implementation. Thus, histograms are represented as multi-dimensional arrays with single-precision, and writing line segments to the arrays is done with an adaptation of Bresenham's line algorithm \cite{bresenham1965algorithm}. Nevertheless, as will be shown next, our implementation outperforms all existing bundling algorithms, even those that make use of the GPU. This being said, a high potential for improving performance even further is left open, for instance, by using off-screen OpenGL rendering for building the histograms and GLSL for smoothing them and for updating the sample points.

The multi-thread implementation of the bundling algorithm has 3 stages: a) resampling; b) histogram building and smoothing; and c) edge advection and smoothing. In a) and c), edges are equally distributed by threads (4 threads were used for tests), while in b) a thread is created for each layer of the histogram. This parallelization strategy has the aim of avoiding variable synchronization and, thus, a thread never writes to a variable that another thread needs to read from or write to. The exception is on stage b) when there is a single bin because only a single thread would be used, waisting CPU resources. Thus, it this case, the line segments of the polylines of all edges are split by the threads because they do not overlap and result in writes to different memory locations within the histogram. Still, when B is 1, histogram smoothing is done by a single thread because the filter is iterative and there are practically no gains in parallelizing this operation. It should be emphasized that the proposed implementation takes full advantage of the multi-thread capabilities of the CPU when using multiple bins, which is the scenario for which the algorithm was primarily designed.

\section{Results and Discussion}
\label{sec5}

All drawings in this paper were constructed with histograms of size 800, {i.e.} the dominant dimension was set to 800 and the smallest dimension was  calculated using the aspect ratio of the bounding box of the graph. All timings of the bundling algorithm exclude graph reading and rendering and were measured in a Macbook Air, Intel Core i5 1.3 GHz, 8GB RAM.  

\begin{figure*}[p]
 \centering
 \includegraphics[width=1\textwidth]{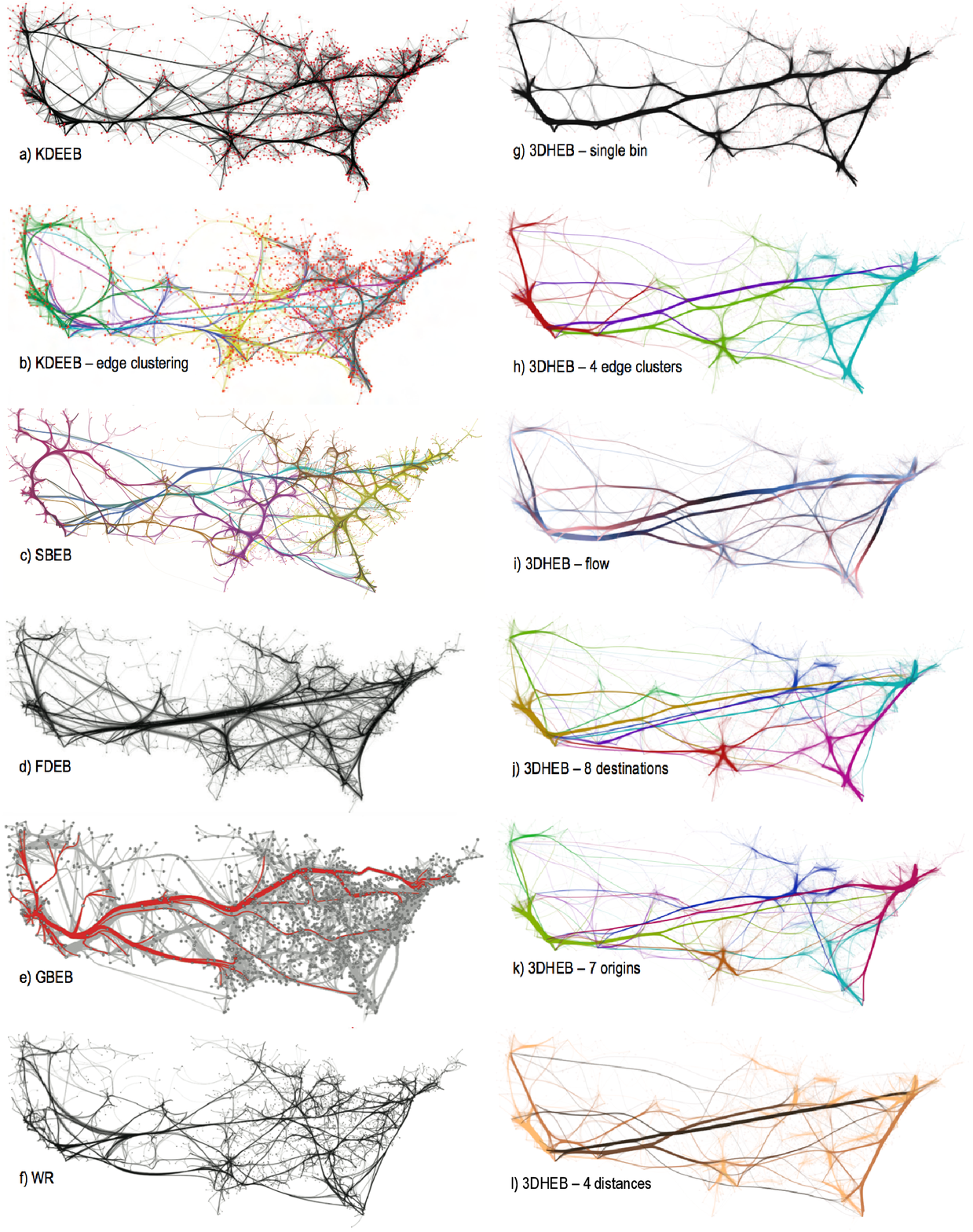}
 \caption{Comparison of bundles using the proposed algorithm (right column) with other bundling algorithms (left column) on the US migration dataset. }
 \label{fig:migration}
\end{figure*}

One of the main features of the proposed technique is the ability to drive bundling using different criteria. This is demonstrated on the US migration dataset (Fig \ref{fig:migration}, right column), where we have bundled the graph with a single bin (Fig \ref{fig:migration}g) as well as with different binning criteria that emphasize different patterns.  Thus, edges were clustered according to the following criteria: origin-destination (Fig \ref{fig:migration}h), orientation  (Fig \ref{fig:migration}i), destination (Fig \ref{fig:migration}j), origin (Fig \ref{fig:migration}k), and distance between origin and destination (Fig \ref{fig:migration}l). Orientation bins were found by discretizing angles in 4 slices centered on $0^\circ$, $90^\circ$, $180^\circ$ and $270^\circ$. For the remaining criteria, bins where found by K-means clustering based on the euclidean distance. The number of bins, $B$, was optimized between 4 and 16 by minimizing the Davies-Bouldin index \cite{db_index}. For each $B$, K-means was executed 100 times with different random initializations. 

\begin{figure*}[htb]
 \centering
 \includegraphics[width=1\textwidth]{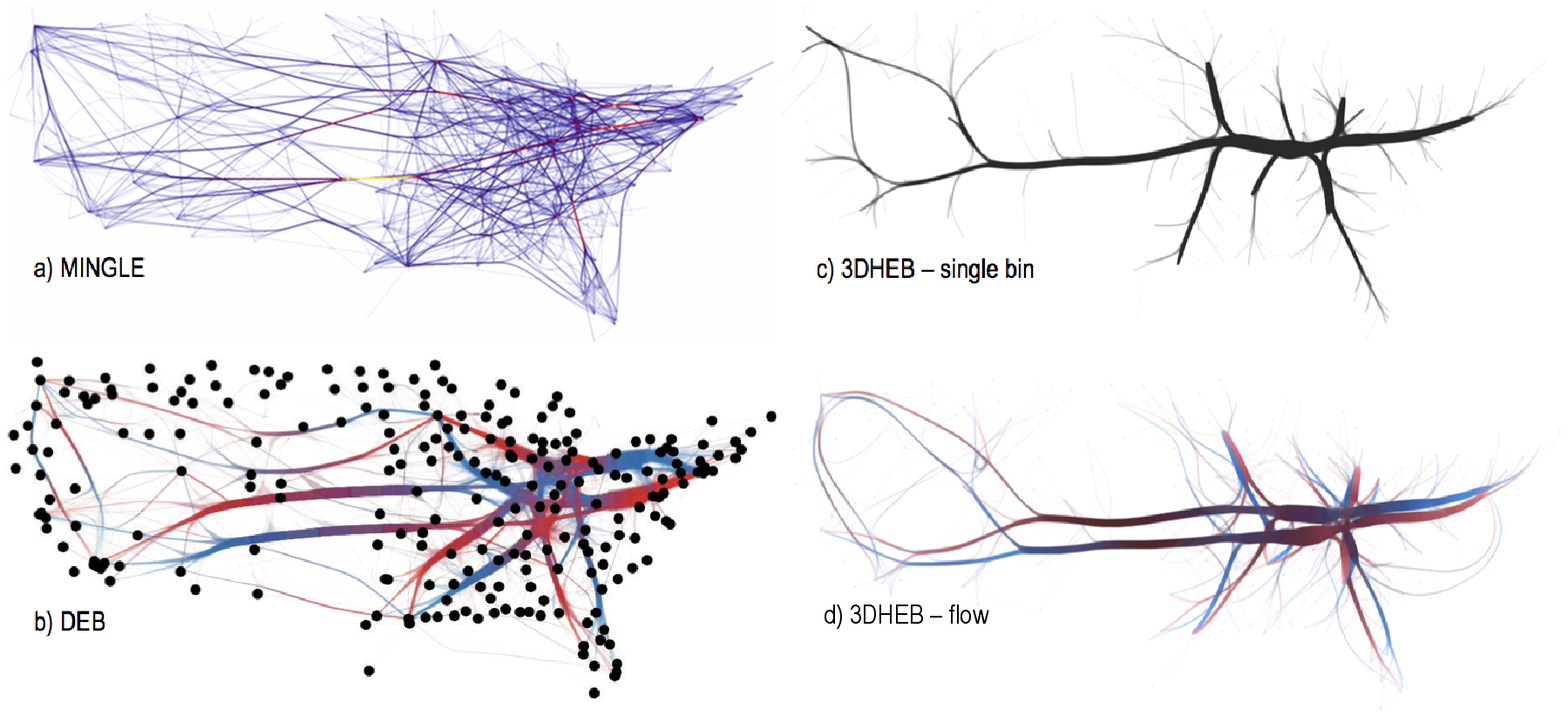}
 \caption{Comparison of bundles using the proposed algorithm (right column) with other bundling algorithms (left column) on the US airlines dataset. }
 \label{fig:airlines}
\end{figure*}

The single-bin bundling shows similar patterns to most of the previous bundling approaches, with less clutter and less over-branching than FDEB, GBEB, SBEB and WR. When comparing to KDEEB, there are several resemblances, as expected, since both methods are based on edge density, and several of the observed differences may be justified by parameters settings, namely in terms of bandwidth and edge smoothing. When using multiple bins, the proposed approach, 3DHEB, shows patterns that none of the previous bundling algorithms is able to reveal without requiring interaction, allowing to immediately answer questions such as, what is the overall flow in the network, where do people come from, where do people go to, what are the longest journeys and how frequent are they? SBEB also uses clustering to group edges but with a fixed criterium for the purpose of determining which edges should be bundled together. In \cite{KDEEB} the authors have used the clusters found by SBEB to independently generate bundled graphs that were then merged together with distinct colors  (Fig \ref{fig:migration}b). Still, such approach fails to answer the previous questions and result in more cluttered bundles than our approach since there is no interaction among edges from different bins, {i.e.} unrelated edges are not repelled. Another advantage of the proposed dataset is that non-spatial properties may be used to guide bundling. This is demonstrated to some extent when using edge distance as the bundling criterium (Fig \ref{fig:migration}l) where it is possible to see edges of the same group  in different locations of the graph without being connected. If, for instance, the age of the people migrating from one place to another is available, it may be used in the same way, and the algorithm would try to bundle together edges that were related by spatial proximity and age group. The US airlines dataset was also used to compare 3DHEB with bundling algorithms that did not present results on the US migration dataset, namely MINGLE and DEB (Fig. \ref{fig:airlines}). It is possible to observe that 3DHEB produces much less cluttered results than MINGLE, while it is able to capture flow in a similar way to DEB with much less computational effort, as will be shown next. 

Performance of the algorithm was evaluated and compared with previously published results for several datasets (Table \ref{tab:times}). The graphs \texttt{net50}, \texttt{net100}, \texttt{net150} and \texttt{pattern1} are large graphs from the University of Florida Sparse Matrix Collection \cite{matrixcoll}. Since the nodes of these graphs do not have locations, the same approach used in \cite{MINGLE} was employed here, consisting in obtaining the coordinates with \texttt{sfdp} from Graphviz \cite{graphviz}, which is an implementation of a force-directed algorithm \cite{sfdp}. The graph \texttt{random} is a randomly generated graph that tries to mimic the one used in \cite{KDEEB} for evaluating the algorithm scalability and has the same number of edges and sample points, being therefore comparable. The \texttt{yeast} graph\footnote{\url{http://github.com/gephi/gephi/wiki/Datasets}} is from the biological domain and was previously used for assessing the performance of MINGLE. Despite we have shown that the proposed approach requires a smaller number of iterations, 10 iterations were used for all datasets, which is the same number of iterations reported in \cite{KDEEB} for the KDEEB benchmarking.

\begin{table*}[tb]
 \caption{Bundling times (in seconds) taken by different edge bundling algorithms. 3DHEB bundling times were measured for 1, 4 and 16 bins. The remaining times were taken from the literature. Algorithms marked with a star use a CPU two times slower than ours, while algorithms marked with a plus also use the GPU, and in the case of KDEEB results are for two distinct GPUs \cite{KDEEB}. The number of sample points used by KDEEB in the airlines and migration dataset is about 20\% higher that ours and it is the same for the random dataset. }
 \label{tab:times}
 \scriptsize
 \begin{center}
   \begin{tabular}{lcc|ccc|cccccccc}
Dataset & Edges & Samples & \multicolumn{3}{c|}{3DHEB} & MINGLE & KDEEB\textsuperscript{+} & SBEB\textsuperscript{+} & WR & GBEB\textsuperscript{*} & OB & FDEB\textsuperscript{*} & DEB\\
&&& 1 & 4 & 16 &&&&&&&&\\ 
\hline
airlines & 1.3K & 66K 		& 		0.4  & 0.7 		& 1.8 	& \textbf{0.1}	& 0.5 - 1.4 	& 6.3 &  & 2.5 	& 3.32	& 19 & 23.8\\
yeast & 6.6K & 69K 			& \textbf{0.5}   & 1.0 		& 2.9 	& 0.9 		&  			&  	&  &  	&		&  & \\
migration & 9.7K & 178K 		& \textbf{0.5}   & 1.0 		& 2.5 	& 1.0 		& 1.5 - 3.6 	& 4.1 & 6.3 &	& 35.75	& 18.8 & 80 \\
random & 200K & 4.80M 		& \textbf{6.1}   & 19.9 	& 72.2 	&  			& 18 - 43 		&  	&  &  	&		&  & \\
net50 & 464K & 3.51M 		& \textbf{5.1}   & 9.3	 	& 29.8 	& 87.1 		&  			& 	 &  &  	&		&  & \\
net100 & 1.00M & 7.25M 		& \textbf{8.4}   & 18.0 	& 52.6 	& 204 		&  			& 	 &  &  	&		&  & \\
net150 & 1.54M & 10.8M 		& \textbf{12.0} & 25.9 	& 68.7 	& 355 		&  			& 	 &  &  	&		&  & \\
pattern1 & 4.65M & 21.6M 	& \textbf{29.4} & 53.7 	& 157 	& 1049 		&  			& 	 &  &  	&		&  & 
   \end{tabular}
 \end{center}
\end{table*}

Since all the algorithms (with the exception of SBEB) do not use binning, a comparison based on the single-bin 3DHEB shows that for all graphs, with the exception of \texttt{airlines}, bundling times of the proposed algorithm are lower. While performance differences tend to be small on smaller graphs, when the number of edges reaches the hundreds of thousands, differences become much larger. In \texttt{pattern1}, the largest graph with near 5 million edges, MINGLE, the previously fastest bundling algorithm, required 1049 seconds while the proposed approach required less than 30 seconds. This is a difference in the order of magnitude that cannot be explained by differences in the hardware. In fact, according to \texttt{cpubenchmark.net} our processor is about 20\% faster than the processor used in \cite{MINGLE}, with our approach being more than 35 times faster in this graph. Even when using 16 bins, 3DHEB remains more than 6 times faster than MINGLE. When considering  KDEEB, comparing computing performance is not so straightforward since the available CPU details are not enough to identify it and because most of the processing is done using GPU. Considering the fastest GPU evaluated by the authors, our CPU implementation of 3DHEB still shows performances 3 times faster on the \texttt{random} graph. Unfortunately the authors did not published results for larger graphs. In smaller graphs, differences tend to be smaller. All the remaining approaches only show timings for the smaller graphs and are all between six and hundreds of times slower than 3DHEB. DEB, the only bundling algorithm that tackled the problem of differentiating flow in networks is tens of times slower in the smaller graph tested here despite being implemented in OpenCL. 

\begin{figure}[htb]
 \centering
 \includegraphics[width=0.9\columnwidth]{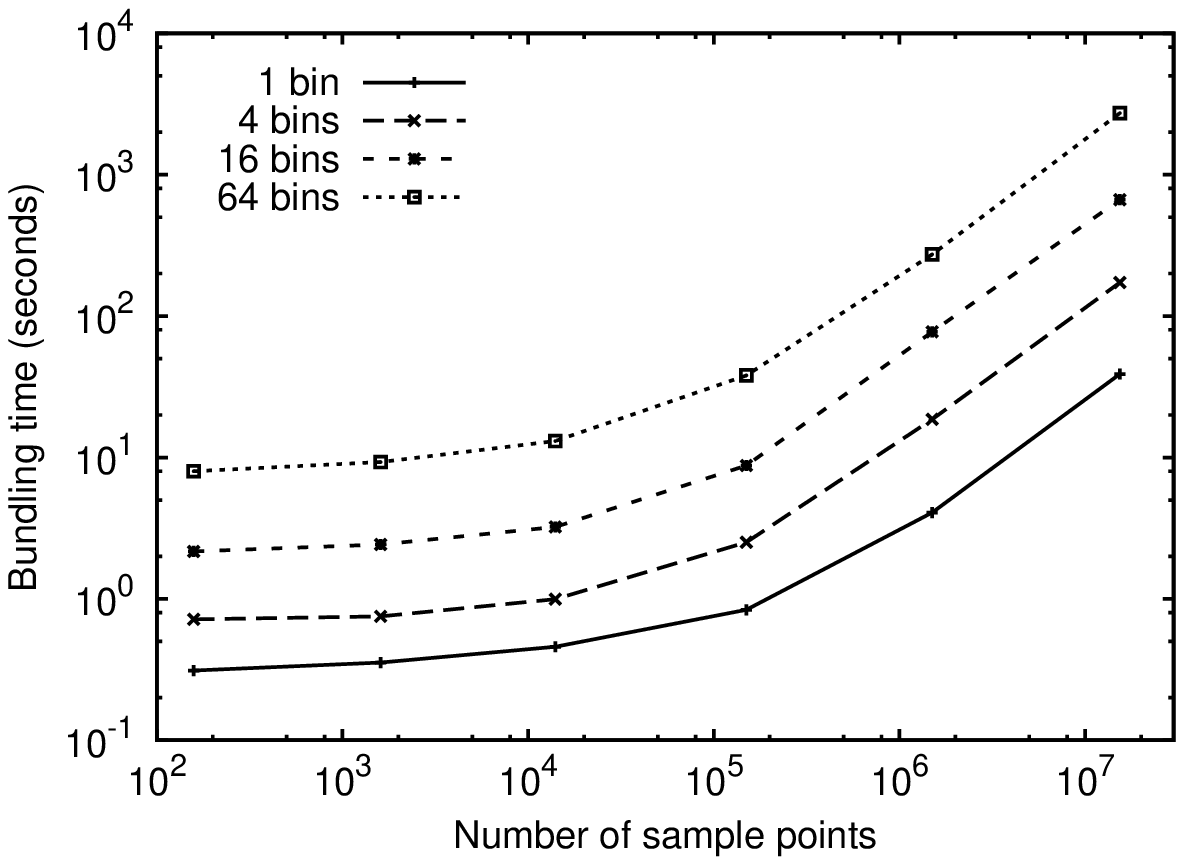}
 \caption{Relation between bundling time and the size of the problem (number of sample points) and number of bins. }
 \label{fig:time_vs_size}
\end{figure}

Overall, results show that the proposed algorithm scales well. The proposed strategy of doing histogram smoothing as a separate step when computing the density map  pays off in larger graphs since it does not depend on the size of the graph, while it becomes a small overhead in small graphs. 
In order to make a deeper assessment of the scalability of 3DHEB, an experiment was conducted where random graphs were bundled with increasing number of edges and with different number of bins. Results (Fig. \ref{fig:time_vs_size}) show that, independently of the number of bins, the relation between the size of the problem and the computing time tend to becomes linear. In smaller graphs, as observed earlier, the smoothing operation creates a small overhead. Results also show, that the algorithm scales well as the number of bins is increased, with computing times increasing proportionally.

\section{Conclusion}
\label{sec6}

This paper presented a new bundling algorithm for general graphs. The proposed approach allows bundling graphs with millions of edges in seconds, outperforming all previous algorithms. In addition, the proposed method allows edge bundling to be driven by user-defined criteria, providing higher insights into the data, higher control in the way that edges are aggregated, and revealing patterns that were hidden or difficult to interpret with previous algorithms.

Regarding future directions, it would be interesting to model more complex relations between groups of edges, especially in graphs that are rich in terms of edge properties, and to study their effects in the generated drawings. The algorithm has shown to coupe with a considerable number of groups (we have tested up to 64 groups), however, as  the number increases, differentiating groups in the drawing becomes a challenge. Interaction techniques may play an important role here, taking advantage of the multiple layers of the histogram and of knowing the distance between layers, which is modeled by the interaction matrix. Finally, a GPU implementation of the algorithm would make it even faster since the algorithm is highly parallelizable and because it makes use of several computer graphics operations that are highly optimized in GPUs.

\acknowledgments{
The author wishes to thank the Future Cities project (European Project FP7, Coordinated Support Action 2012-316296) for supporting this work. }

\bibliographystyle{abbrv}
\bibliography{bundling}
\end{document}